\documentclass[11pt]{article}
\usepackage[utf8]{inputenc}
\pdfoutput=1
\usepackage{amssymb,amsmath,mathrsfs,enumerate}
\usepackage{graphicx,rotate,multicol}
\usepackage{float}
\usepackage{tocloft}
\usepackage{subfig}
\usepackage[margin=10pt,labelfont=bf]{caption}
\usepackage{cite}
\usepackage{soul}
\usepackage[colorlinks=true,
            linkcolor=blue,
            urlcolor=blue,
            citecolor=teal]{hyperref}
\usepackage{multirow}

\makeatletter
\let\Hy@linktoc\Hy@linktoc@page
\makeatother

\usepackage{color}
\definecolor{ourcolor}{rgb}{0.7, 0.25, 0.05}
\usepackage{tikz,braket}
\long\def\rpl#1!!#2!!{\textcolor{red}{#1} \textcolor{blue}{#2}}

\let\tilde=\widetilde

\let\bar=\overline

\def \order(#1){{\mathcal O} \left(#1 \right)}

\evensidemargin=\oddsidemargin
\allowdisplaybreaks
\begin{document}

\begin{titlepage}

\vspace*{-5mm}

\centering{\LARGE{ \bf Detecting Bosonic Dark Matter with Neutron Stars}}

\centering 

{\bf Tarak Nath Maity,} $^{a,b}$  \footnote{tarak.maity.physics@gmail.com}
{\bf Farinaldo S Queiroz} $^{c,d,e}$  \footnote{farinaldo.queiroz@ufrn.br}\\

\vspace{3mm}
$^{a}$\,{\it Centre  for  High  Energy  Physics,  Indian  Institute  of  Science,  Bangalore  560012,  India} \\
\vspace{1.5mm}
$^{b}$\,{\it Institute of Physics, Sachivalaya Marg, Bhubaneswar 751005, India} \\
\vspace{1.5mm}
 $^c$\,{\it International Institute of Physics, Universidade Federal do Rio Grande do Norte,  Campus Universitario, Lagoa Nova, Natal-RN 59078-970, Brazil}\\
 \vspace{1.5mm}
 $^d$\,{\it Departamento de F\'isica, Universidade Federal do Rio Grande do Norte, 59078-970, Natal, RN, Brasil}\\
  \vspace{1.5mm}
 $^e$\, {\it Millennium Institute for Subatomic Physics at the High Energy Frontier (SAPHIR) Chile}

\date{}


\begin{abstract}
What if the dark matter-nucleon scattering cross section is too small to be detected by direct detection experiments? It is well known in the literature that some interactions lead to dark matter-nucleon scattering cross sections that can be velocity and momentum suppressed. We show that in the case of bosonic dark matter, neutron star spectroscopy offers a possible detection. Firstly, we discuss the case of scalar dark matter with scalar, pseudoscalar, and vector mediators. Later, we do this exercise for vector dark matter. We show that, depending on the nature of dark matter and the interaction involved, neutron stars can improve the sensitivity on the dark matter-nucleon scattering cross section by orders of magnitude, representing a major step forward in the dark matter siege.  
\end{abstract}

\end{titlepage}

\newpage

\hrule \hrule
\tableofcontents
\vskip 10pt
\hrule \hrule 

\section{Introduction}
\label{sec:intro}
Several independent astrophysical and cosmological observations ranging from galactic to extragalactic length scale firmly attest that a major component of matter in the Universe consists of dark matter (DM) \cite{Jungman:1995df, Bertone:2010zza}. However, this observational evidence for DM is gravitational in nature. Thus, its nature is yet unknown \cite{Jungman:1995df, Bertone:2010zza, Arcadi:2017kky}. An attractive possibility is that the DM may be comprised of elementary particles, and possesses  nongravitational interactions with the Standard Model (SM) fields. Consequently, a lot of effort has been made to detect the interactions of DM with the visible sector, namely direct detection, indirect detection, colliders, etc. We briefly review these searches and outline later the complementary aspect that neutron stars offer concerning DM searches.

The Earth-bound direct detection experiments rely on the assumption that the DM may scatter off of the nuclei of the target material, and hoping that the recoil energy is above the energy threshold of the detector, would allow a possible identification of the DM interactions \cite{Goodman:1984dc,Drukier:1986tm}. It should be noted that while the detection of DM can happen in this case, its fundamental properties such as spin, and $CP$ nature are still to be unveiled \cite{Queiroz:2016sxf,Queiroz:2018utk}. Notably, direct detection experiments have made excellent progress in the last few decades, though no conclusive signal was found thus far \cite{Akerib:2016lao, Cui:2017nnn, Aprile:2018dbl}. Direct detection experiments are severely limited by their energy threshold, which is relatively high for DM-nucleon scattering \cite{Abdalla:2019xka}. Moreover, the presence of the atmospheric and solar neutrino flux, the so-called neutrino background, potentially  limits the sensitivity of these experiments \cite{Monroe:2007xp, Vergados:2008jp, Strigari:2009bq, Gutlein:2010tq}.

Regarding indirect detection of DM, it depends on the DM annihilation cross section into either SM particles \cite{CTAConsortium:2018tzg} or some secluded sector \cite{Pospelov:2007mp,Pospelov:2008jd,Fortes:2015qka,Profumo:2017obk,Siqueira:2019wdg}. Either way, a potential signal from DM is subject to astrophysical uncertainties concerning the DM density in the galactic center or dwarf spheroidal galaxies, see for instance \cite{Gaskins:2016cha,Strigari:2018utn}. Claiming solidly the discovery of DM would need the observation of signals at different targets. The search for DM using neutrino telescopes can be twofold. It can be based on the neutrino signal rising from PeV scale DM \cite{Murase:2015gea, Capanema:2020oet,Capanema:2020rjj,Bhattacharya:2019ucd}, or from looking for neutrinos coming from particular celestial objects such as the Sun. In the latter, the signal is governed by the DM-nucleon scattering cross section assuming the DM particles thermalize in the Sun \cite{Leane:2017vag, ANTARES:2019svn,Aartsen:2020tdl}. However, one can still probe DM interactions, relaxing the assumption of thermalization \cite{Fornengo:2017lax}. 

Concerning collider searches, a potential signal from DM is typically interpreted as missing energy and should be in most cases accompanied by some visible particle, used as a detector trigger \cite{Abercrombie:2015wmb}. The Large Hadron Collider (LHC) is not sensitive to either light $1$~GeV or heavy ($> 1$~TeV scale) DM, and in several simplified models, it is more promising to probe under-abundant DM, which is tied to large couplings to fermions \cite{Kahlhoefer:2017dnp}. Furthermore, the LHC can not claim the discovery of DM, but rather, the observation of a sufficiently long-lived particle. 

Given these shortcomings present in indirect, direct detection, and collider searches, the use of celestial bodies as laboratories to probe for DM interactions surfaced \cite{Gould:1987ju, Gould:1987ir, Press:1985ug, Goldman:1989nd}, and have been explored widely \cite{Kouvaris:2007ay,   Guver:2012ba,  Busoni:2013kaa, Cermeno:2015efs, Cermeno:2016olb, Cermeno:2017ejm, Cermeno:2018qgu, Bell:2018pkk, Camargo:2019wou, Bell:2019pyc, Garani:2019fpa, Acevedo:2019agu, Dasgupta:2020dik, Garani:2020wge, Leane:2020wob}.\footnote{ We highlight that our study does not include the case of asymmetric DM, the capture of which can lead to formation of black hole \cite{deLavallaz:2010wp, McDermott:2011jp, Kouvaris:2010jy, Bell:2013xk ,Garani:2018kkd, Dasgupta:2020mqg}.} When a compact object passes through the DM halo, the DM may lose its energy sufficiently through its scattering with SM states, which resides inside the compact object and subsequently get captured by the compact object. This DM capture may lead to considerable heating of compact objects like neutron stars. This heating could be detectable by the next-generation infrared telescopes such as the James Webb Space Telescope (JWST). Objects like neutron stars are good laboratories for DM for several reasons. Firstly, the high density of SM states in the neutron stars facilitates the DM capture process. Further, in the vicinity of neutron stars, DM particles move at velocities much higher than the average halo velocity ($v \sim 10^{-3}c$), making relativistic effects important, conversely  to direct detection experiments \cite{Camargo:2019wou,Bell:2020jou}. It is important to emphasize that the detection of DM using neutron stars suffers from large systematic uncertainties, as the equation of state for neutron stars in the presence of DM with elevated temperatures are poorly understood. Therefore, a future detection of DM based on neutron stars relies on the observation of a population of neutron stars with temperatures higher than expected from theoretical predictions. In this case, the equation of state that these neutron stars have, which might be different, has a little effect on the overall conclusions. Hence, neutron stars can indeed offer a concrete and exciting opportunity in the DM siege. On the other hand, if a DM signal is observed in a direct detection experiment, for instance, and later an observed heat in neutron stars is observed, neutron stars could certainly strengthen the case for a DM detection.

We have seen that direct detection, indirect detection, and collider probes have their pros and cons concerning the DM search. Neutron stars are not different, but bring new ingredients: (i) they can overcome the low energy threshold present in direct detection experiments, allowing us to probe light DM; (ii) they may test the DM-nucleon scattering cross section down to much lower values compared to current and future neutrino telescopes; (iii) despite relying on an astrophysical observation, neutron stars are sensitive to the DM-neutron scattering cross section differently from indirect detection; (iv) they probe the mass range $1{\, \rm MeV}-100 {\,\rm TeV}$, which extends collider searches. Hence, neutron stars indeed offer an important and orthogonal avenue to be explored in the future. Therefore, if a DM signal is observed by any of the standard observations, neutron stars will certainly help to discriminate models and unveil the nature of DM.

From the point of view of the DM models, the neutron stars have mostly been studied in the context of fermionic DM within the effective field theoretical (EFT) \cite{Bell:2018pkk, Bell:2019pyc} framework.\footnote{Note that the case of scalar DM has been explored in effective field theoretical framework in references \cite{Joglekar:2020liw}.} In this paper, we extend these analyses by exploring both the scalar and vector DM scenarios. In this work, we have considered some well-known simplified Lagrangian having both scalar and vector mediators.  We assess the neutron star sensitivity to the DM-nucleon scattering cross section for two different temperature measurements, and compare our findings with those stemming from direct detection experiments. 

The rest of the paper is organized as follows. In section \ref{sec:crate}, we briefly review the capture framework for a neutron star. In section \ref{sec:sdm} and \ref{sec:vdm}, we have provided our results with simplified spin-$0$ and spin-$1$ DM Lagrangian respectively. Finally, we conclude in section \ref{sec:conclusion}.

\section{Capture Rate}
\label{sec:crate}
In the DM halo, particles move with nonrelativistic velocities, but they may get accelerated up to a speed of $\mathcal{O}(0.3 c)$, when they encounter the steep gravitational potential of a massive compact object like a neutron star. After falling into the gravitational potential of the neutrons star, the DM particles can lose sufficient energy through interactions with nucleons in the neutron star and then get trapped inside the neutron star. Within a relevant timescale, DM capture contributes maximally to the neutron star heating. In other words, the entire initial kinetic energy of DM contributes to the heating of the neutron star. This contribution can be written as \cite{Raj:2017wrv, Bell:2018pkk}
\begin{equation}
T= 1700\, {\rm K}\, f^{1/4} \left( \frac{\rho_{\chi}}{0.4\, {\rm GeV/cm}^3} \right) \left(\frac{\rm {Erf}(x)}{x\, \rm {Erf}(1)}\right)^{1/4},
\label{eq:temp}
\end{equation}
where the DM density is represented by $\rho_{\chi}$. Although this can be as large as $\sim 10^3 \,{\rm GeV/cm^3}$ \cite{McCullough:2010ai},  to obtain a conservative limit we fixed it to $0.4 \,{\rm GeV/cm^3}$. In equation \eqref{eq:temp} $x=v_{\star}/(230 \, \rm {km/s})$, $v_{\star}$ is the neutron star velocity, which it is assumed to be the velocity of the Sun \cite{Bell:2018pkk}.  The capture efficiency $f$ is defined as
\begin{equation}
f \sim {{\rm Min} \left[ \frac{\sigma_{\chi n}}{\sigma_{\rm th}},1\right]},
\label{eq:f}
\end{equation}
where $\sigma_{\chi n}$ denotes the DM-neutron scattering cross section, and $\sigma_{\rm th}$ the threshold cross section. For a neutron star of mass $M$ and radius $R$, a reasonable estimate of $\sigma_{\rm th}$ is found using the following equation:
\begin{equation}
\sigma_{\rm th} = \pi R^2 \frac{M}{m_n},
\end{equation}
where $m_n$ is the mass of the neutron. To obtain the constraint on the relevant parameter space, we have chosen a benchmark neutron star of radius $R=10\,$km and mass $M=1.5 M_{\odot}$.  It has been argued in reference \cite{baryakhtar:2017dbj} that upcoming infrared telescopes like JWST \cite{Gardner:2006ky}, Thirty Meter Telescope (TMT) \cite{Crampton:2008gx} and European Extremely Large Telescope (E-ELT) \cite{Maiolino:2013bsa} have the potential to measure the heat in neutron stars coming from DM interactions. In this work, we will focus on the DM mass range $1 \, \rm{GeV} \leq m_{\chi} \leq 10^{6} \, \rm{GeV}$. We highlight that for DM mass less than $1$~GeV, Pauli blocking becomes important \cite{Bell:2020jou, Bell:2020lmm, Joglekar:2019vzy, Joglekar:2020liw}, whereas for DM mass greater than $10^6$ GeV, multiple collisions are required for the DM to lose enough kinetic energy \cite{Gould:1991va, Albuquerque:2000rk, Mack:2007xj, Bramante:2017xlb, Dasgupta:2019juq, Ilie:2020vec}. These effects are not important for the mass range explored here. However, we caution the reader about equation \eqref{eq:temp} which features simplifying assumptions. For instance, it is assumed that the neutron star interior is nonrelativistic in nature, which is consistent for DM-nucleon scattering considered in this work \cite{Bell:2020jou}.  For other targets like electron and muons, relativistic effects will have a considerable impact on the capture rate \cite{Bell:2020lmm, Joglekar:2019vzy, Joglekar:2020liw}. Furthermore, the equation of state of the neutron star is poorly understood in the presence of DM and temperatures much larger than zero \cite{Raithel:2019gws,Bhat:2019tnz,Calder:2020kqs,Sen:2020edi,Ghosh:2021bvw}. Moreover, the thermodynamics of a neutron star is rather complex \cite{Hartle:1968si,Thorne:1997kt,Page:2004wb, Page:2012se, Page:2013uwa, Wijnands:2017jsc}.

That said, our reasoning goes as follows: we select a simplified DM model, compute the DM-neutron scattering cross section, plug in this cross section into equation \eqref{eq:f}, and then obtain the neutron star temperature using equation \eqref{eq:temp}. Having described the procedure, we now discuss the simplified DM models.

\section{Scalar Dark Matter}
\label{sec:sdm}
In this section, we consider a spin-$0$ DM interacting with nucleons in the context of neutron star heating. This has been studied in the context of EFT framework, however, in the EFT framework one may lose important information \cite{Goodman:2011jq,Shoemaker:2011vi, Buchmueller:2013dya,DeSimone:2016fbz}. We will use simplified DM models because they keep a closer contact with the ultraviolet completions with a manageable number of parameters, namely coupling constants, the masses of particles \cite{Arcadi:2017kky}. We will explore a set of simplified Lagrangians to represent the interaction between the DM and SM sectors.

Firstly, we consider a complex scalar DM ($\chi$) interacting with the visible sector through a scalar ($A$) or vector ($A^{\mu}$) mediator. The corresponding simplified Lagrangian can be written as \cite{Berlin:2014tja}
\begin{subequations}
\begin{align}
-\mathcal{L}^S_1  &\supset    \mu_{\chi} \chi^{\dagger}  \chi A + \lambda_{ fs}\, \bar{f} f  A \label{subeq:LS1}\\
-\mathcal{L}^S_2  &\supset    \mu_{\chi} \chi^{\dagger}  \chi A + \lambda_{ fp}\, \bar{f} i \gamma_5 f  A \label{subeq:LS2}\\
-\mathcal{L}^S_3  &\supset    i g_{\chi}  (\chi^{\dagger} \partial_{\mu} \chi-\chi \partial_{\mu} \chi^{\dagger}) A^{\mu} + g_{ fv}\, \bar{f} \gamma_{\mu} f  A^{\mu}  \label{subeq:LS3}\\
-\mathcal{L}^S_4  &\supset  i g_{\chi}  (\chi^{\dagger} \partial_{\mu} \chi-\chi \partial_{\mu} \chi^{\dagger})A^{\mu} + g_{ fa}\, \bar{f} \gamma_{\mu} \gamma^5 f  A^{\mu}, \label{subeq:LS4}
\end{align}
\label{eq:scalarL}
\end{subequations}
where the $f$ index represents the SM quarks, and $\mu_\chi$, $\lambda_f$ and $g_f,g_\chi$ are couplings constants. Notice that $\mu_\chi$ is the dimensional couplings. Throughout the paper, the dimensional coupling $\mu_{\chi}$ has been fixed to  $\lambda_{\chi} m_{\chi}$. The numerical value of the all the relevant dimensionless couplings have also been fixed to $0.1$, i.e. $g_i=\mu_{\chi}/m_{\chi}=\lambda_i=0.1$.
\begin{figure*}[t]
\begin{center}
\subfloat[\label{sf:S1}]{\includegraphics[scale=0.26]{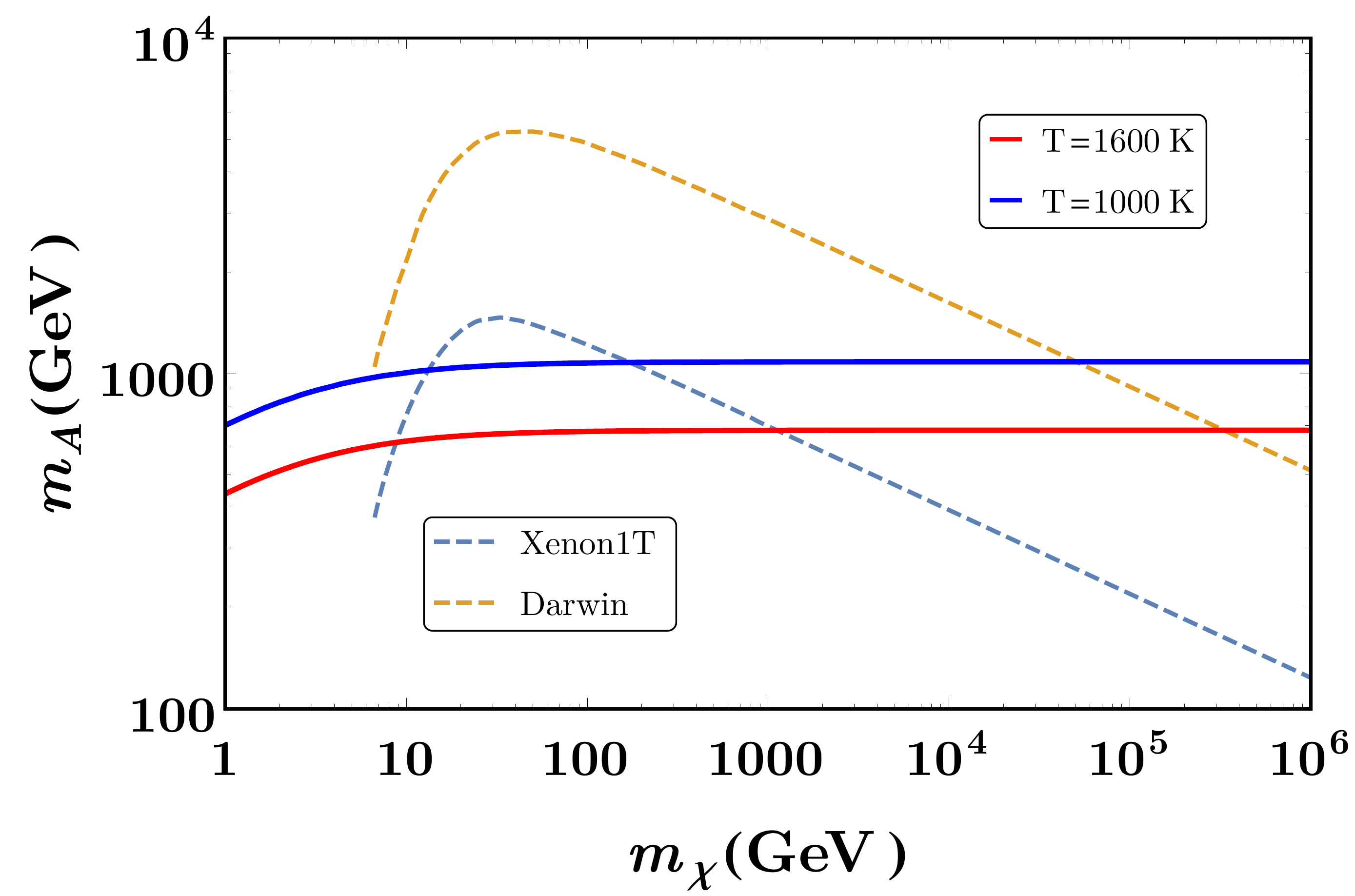}}~~~
\subfloat[\label{sf:S2}]{\includegraphics[scale=0.26]{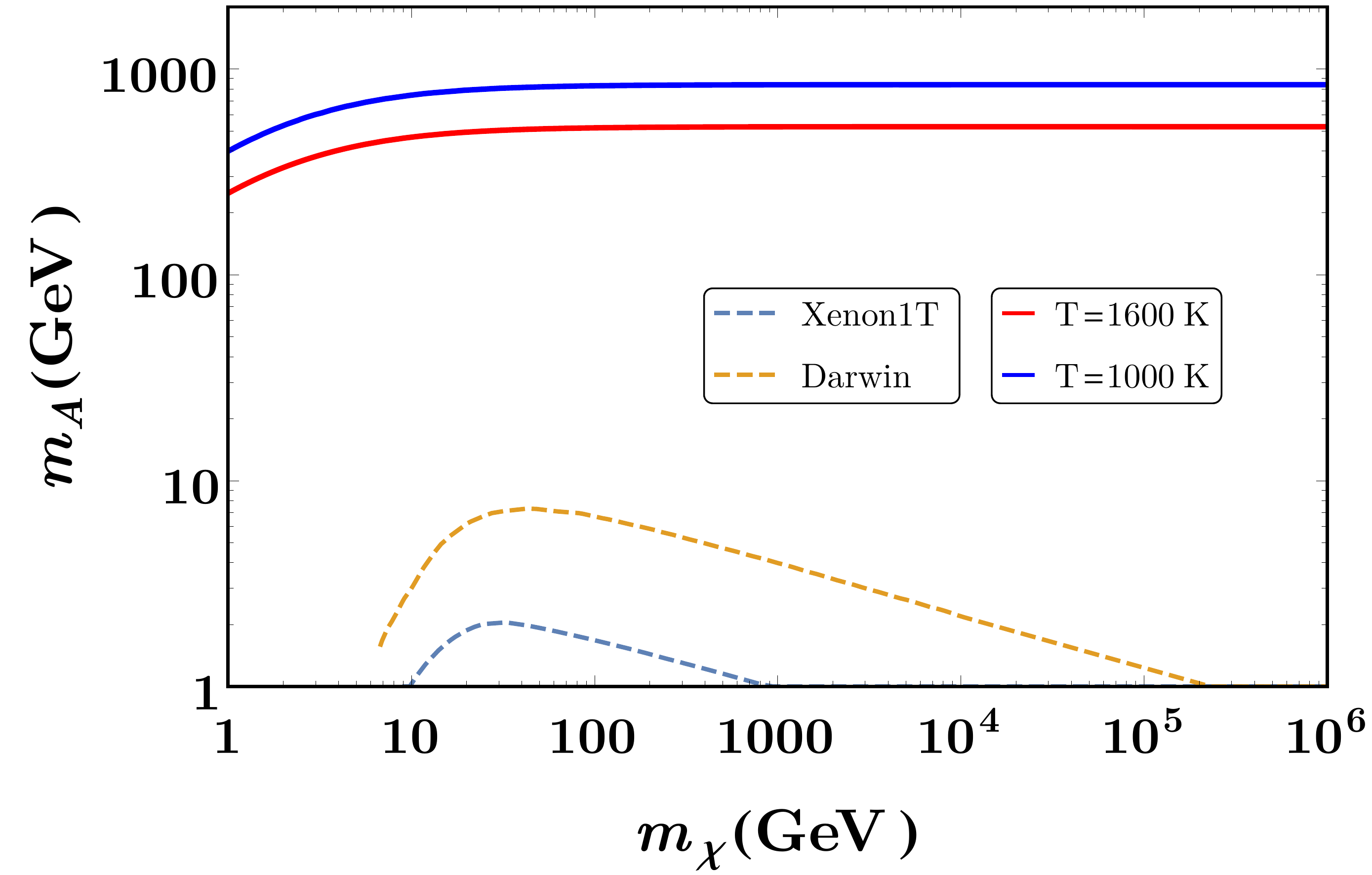}}\\
\subfloat[\label{sf:S3}]{\includegraphics[scale=0.26]{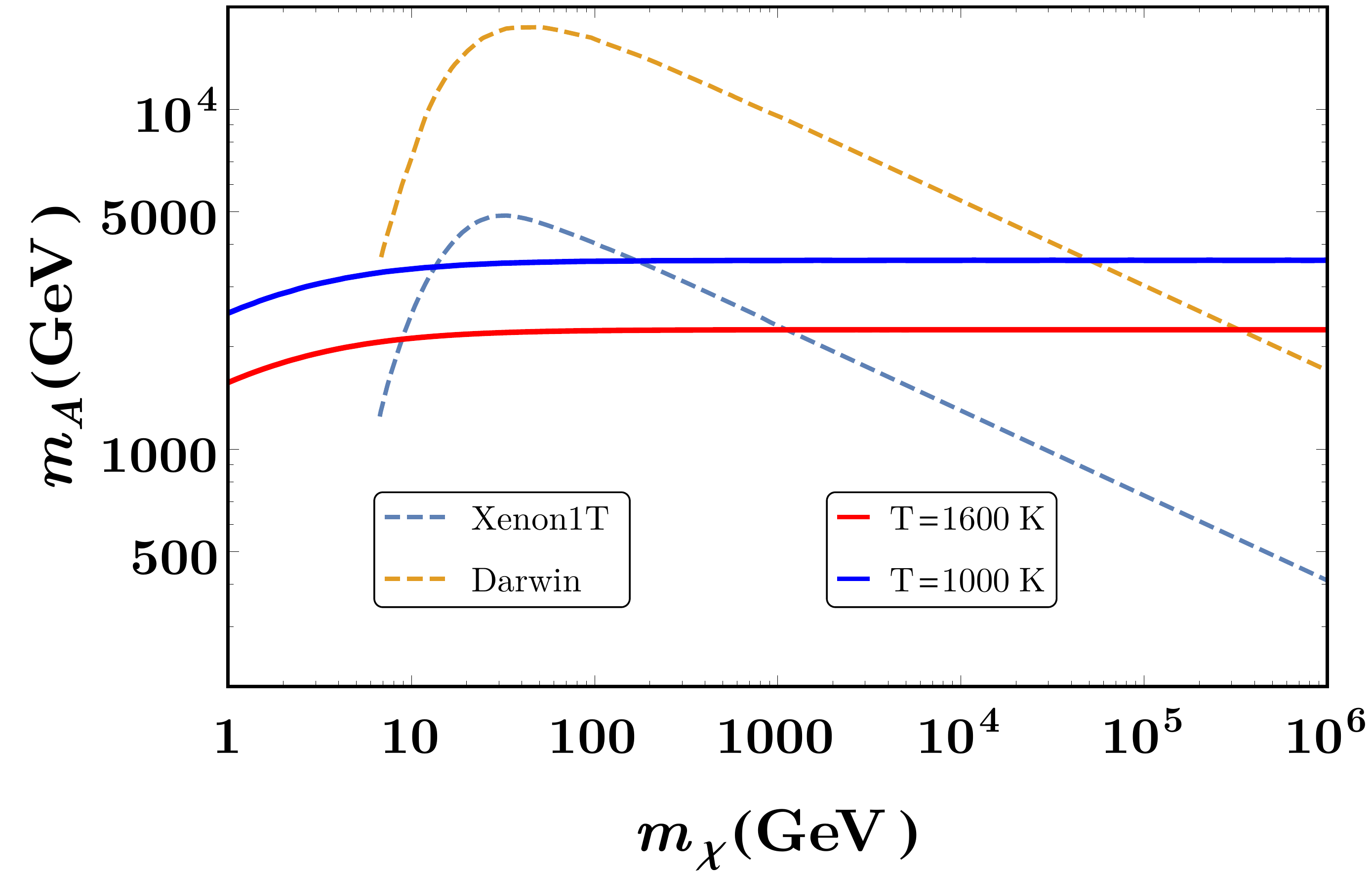}}~~~
\subfloat[\label{sf:S4}]{\includegraphics[scale=0.26]{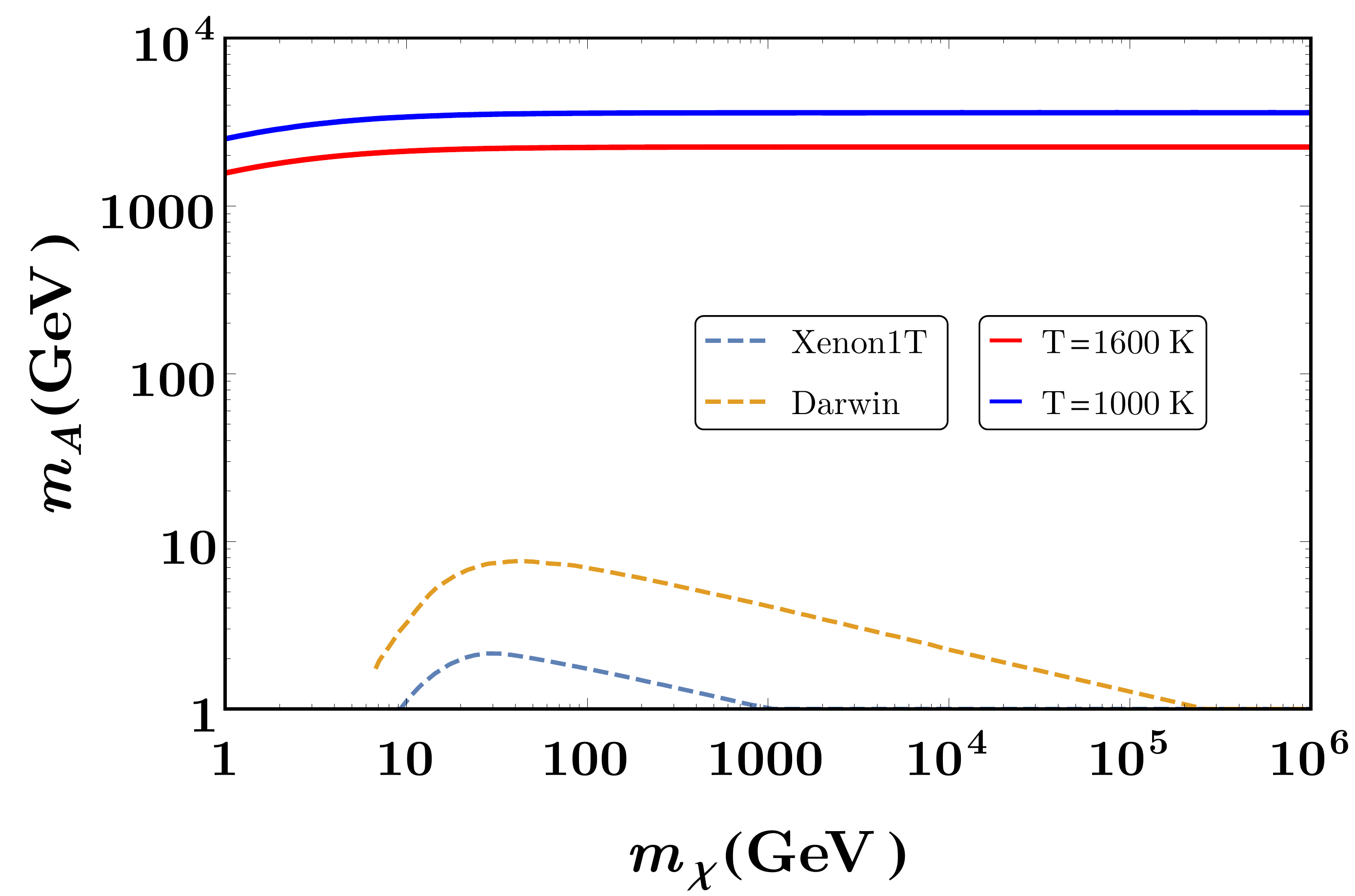}}
\caption{Limits on $m_{\chi}-m_A$ plane from neutron star heating through DM-neutron elastic scattering. In each panel ,the red and blue contours correspond to neutron star temperatures $1600\,$ and $1000\,$K, respectively. The complementary bounds from Earth-based direct detection experiments Xenon1T  and Darwin  are shown by dashed light blue and brown lines. We take $g_i=\mu_{\chi}/m_{\chi}=\lambda_i=0.1$. (a) For the interaction, Lagrangian is given in equation \eqref{subeq:LS1}. (b) For the interaction given in equation \eqref{subeq:LS2}. (c) For vector current interaction given in equation \eqref{subeq:LS3}. (c) For axial current interaction given in equation \eqref{subeq:LS4}.}
\label{fig:NS-DD-S}
\end{center}
\end{figure*}
The associated differential DM-nucleon cross sections are 
\begin{subequations}
\begin{align}
\frac{d\sigma^S_1}{d\cos \theta} & =  \frac{1}{32 \pi s} \frac{\mu_{\chi}^2 \, \tilde{f}_n^2 }{\left(t-m_A^2\right)^2+(\Gamma_{A,1}^S m_A )^2 } \left(4 m_n^2 - t \right)\\
\frac{d\sigma^S_2}{d\cos \theta} &=  \frac{1}{32 \pi s} \frac{\mu_{\chi}^2 \, \tilde{t}_n^2 }{\left(t-m_A^2\right)^2+(\Gamma_{A,2}^S m_A )^2} \left(- t \right) \\
\frac{d\sigma^S_3}{d\cos \theta} & =  \frac{1}{32 \pi s} \frac{g_{\chi}^2 \, \tilde{b}_n^2 }{\left(t-m_A^2\right)^2+(\Gamma_{A,3}^S m_A )^2}\left( -2 s(m_{\chi}^2+m_n^2)+s^2+ st + (m_{\chi}^2+m_n^2)^2 -t m_n^2\right) \\
\frac{d\sigma^S_4}{d\cos \theta} & =  \frac{1}{32 \pi s} \frac{g_{\chi}^2 \, \tilde{a}_n^2 }{\left(t-m_A^2\right)^2+(\Gamma_{A,4}^S m_A )^2}\left( -2 s(m_{\chi}^2+m_n^2)+s^2+ st + (m_{\chi}^2-m_n^2)^2 \right),
\end{align}
\label{eq:sxsection}
\end{subequations}
where $s$ and $t$ are the Mandelstam variables, $m_{\chi}$ the DM mass, and $m_A$ and $\Gamma_{A,i}^S$ the mass and decay width of the mediator. The expression for the decay width $\Gamma_{A,i}^S$ can be evaluated using equation \eqref{eq:scalarL}, and it can be found in \cite{Berlin:2014tja, Arcadi:2017kky}.  We have summed over the relevant quark operators to obtain the DM-nucleon scattering cross section above. The corresponding nucleon level couplings are denoted by a tilde. They are related to quark levels coupling through the following relations:
\begin{subequations}
\begin{align}
\tilde{f}_n &= \lambda_{fs}\, \frac{m_n}{\rm GeV} \left(\frac{7}{9} \sum_{q=u,d,s}f^n_{T_q} +\frac{2}{9} \right)  \\
\tilde{t}_n &= \lambda_{fp} \, \frac{m_n}{\rm GeV}  \sum_{q=u,d,s}f^{(5n)}_{q} \\
\tilde{b}_n &= 3 \lambda_{fv} \\
\tilde{a}_n &=  \lambda_{fa} \sum_{q=u,d,s}\Delta^{(n)}_{q} 
\end{align}
\label{eq:formfactors}
\end{subequations}
We adopted $f^n_{T_u}=0.026, ~f^n_{T_d}=0.02, ~f^n_{T_s}=0.043$ \cite{Junnarkar:2013ac}, and $f^{5n}_{u}=-0.42, ~f^{5n}_{d}=0.85, ~f^{5n}_{u}=-0.08$, with $\Delta^n_u=0.84$, $\Delta^n_d=-0.43$, $\Delta^n_s=-0.09$ \cite{Cheng:2012qr}. With these results at hand, we use equation \eqref{eq:sxsection} to calculate the heat in the neutron star through equation \eqref{eq:temp}. Having in mind the temperature measurements of neutron stars,  we derive our limits accordingly. We put our results into perspective by comparing our findings with the limits stemming from direct detection experiments, which are displayed in figure \ref{fig:NS-DD-S}.

In figure \ref{fig:NS-DD-S}, the red and blue solid lines correspond to a neutron star heating temperature  of $1600\,$ and $1000\,$K, respectively. The light blue and yellow dashed lines correspond to limits from Earth-based direct detection experiments Xenon1T \cite{Aprile:2018dbl, Aprile:2019dbj} and Darwin \cite{Aalbers:2016jon}, respectively. The limits from these direct detections have been calculated using the nonrelativistic version of the differential cross section given in equation \eqref{eq:sxsection}.\footnote{The direct detection limits on various non-relativistic effective DM nucleon operators has extensively been studied in \cite{Fan:2010gt, Fitzpatrick:2012ix, Fitzpatrick:2012ib, DelNobile:2013sia, Crivellin:2014qxa, DEramo:2014nmf, DEramo:2016gos, Catena:2019hzw}.} In the nonrelativistic limit, both the scalar mediated interaction and spin-1 mediated interaction with DM and quark vector current is reduced to contact interaction  \cite{DelNobile:2013sia}. The direct detection rates of such interactions are coherently enhanced by the number of nucleons present in the target, resulting in stringent constraints on the mediator mass. This has been exhibited in both the upper and lower left panels of figure \ref{fig:NS-DD-S}. Setting aside XENON1T, we highlight that curves in the plots represent the potential exclusion regions of these probes. In the upper left panel, DARWIN will be able to exclude mediator masses up to 6 TeV, whereas neutron stars will be able to probe mediator masses up to 1 TeV. Although, the bounds from neutron stars can be more stringent than current and future direct detection experiments in light and heavy DM mass regions. Interestingly, even considering the ultimate DARWIN sensitivity, we notice that for DM masses above $40$~TeV and below $10$~GeV, neutron stars constitute a more promising probe. For the largest part of the parameter space, DARWIN will give rise to stronger bounds.

On the other hand, for the simplified models described in equation \eqref{subeq:LS2} and equation \eqref{subeq:LS4}, which result in a spin-dependent and velocity suppressed scattering cross section, thus the sensitivity of direct detection experiments is weakened. Hence, neutron star probes can be notoriously more constraining than direct detection experiments. This has been demonstrated in both the upper and lower right panel of figure \ref{fig:NS-DD-S}.\footnote{It should be noted that for axial-vector quark coupling, inclusion of loop effects may improve these direct detection limits by two orders of magnitude \cite{Crivellin:2014qxa, DEramo:2014nmf, DEramo:2016gos}, rendering neutron stars still a better laboratory for these DM interactions.} We conclude that because of the mass range of interest, measurements of the neutron star temperatures will be paramount to the detection of DM.

%
%
\section{Vector Dark Matter}
\label{sec:vdm}

 Vector DM arises in several well-motivated Abelian and non-Abelian gauge theories \cite{Hambye:2008bq, Davoudiasl:2013jma, Nomura:2020zlm, Ramos:2021omo}. Vector DM might interact with the SM particles via scalar fields or vector mediators. Vector DM is phenomenologically appealing because the interaction strength with SM particles is typically not suppressed if one wants to reproduce the DM relic density. This allows one to easily test them using data from direct, indirect, and collider experiments. Instead of going through every single vector DM model, we will represent vector DM models through the relevant interactions written as \cite{Berlin:2014tja}
\begin{subequations}
\begin{align}
-\mathcal{L}^V_1  &\supset    \mu_{\chi}\, \chi_{\mu}^{\dagger}\, \chi^{\mu} \,  A + \lambda_{ fs}\, \bar{f} f  A \label{subeq:LV1}\\
-\mathcal{L}^V_2  &\supset    \mu_{\chi}\, \chi_{\mu}^{\dagger}\, \chi^{\mu} \,  A  + \lambda_{ fp}\, \bar{f} i \gamma_5 f  A \label{subeq:LV2}\\
-\mathcal{L}^V_3  &\supset  i g_{\chi} \left(\chi_{\nu}^{\dagger} \partial_{\mu} \chi^{\nu} - \chi_{\nu} (\partial_{\mu} \chi^{\dagger \nu} ) \right)A^{\mu}+ \lambda_{ fv}\, \bar{f} \gamma_{\mu} f  A^{\mu} \label{subeq:LV3}\\
-\mathcal{L}^V_4  &\supset   i g_{\chi} \left(\chi_{\nu}^{\dagger} \partial_{\mu} \chi^{\nu} - \chi_{\nu} (\partial_{\mu} \chi^{\dagger \nu} ) \right)A^{\mu} + \lambda_{ fa}\, \bar{f} \gamma_{\mu} \gamma^5 f  A^{\mu} \label{subeq:LV4}
\end{align}
\label{eq:vectorL}
\end{subequations}
 Certainly, this is not a complete list, but it covers several relevant scenarios of vector DM. The corresponding differential DM-nucleon scattering cross sections are given by
\begin{subequations}
\begin{align}
\frac{d\sigma^V_1}{d\cos \theta} & =  \frac{1}{32 \pi s} \frac{\mu_{\chi}^2 \, \tilde{f}_n^2 }{\left(t-m_A^2\right)^2+(\Gamma_{A,1}^V m_A )^2} \frac{12 m_{\chi}^4-4 m_{\chi}^2 t +t^2}{12 m_{\chi}^4} \left(4 m_n^2 - t \right)\\
\frac{d\sigma^V_2}{d\cos \theta} &=  \frac{1}{32 \pi s} \frac{\mu_{\chi}^2 \, \tilde{t}_n^2 }{\left(t-m_A^2\right)^2+(\Gamma_{A,2}^V m_A )^2} \frac{12 m_{\chi}^4-4 m_{\chi}^2 t +t^2}{12 m_{\chi}^4} \left(- t \right) \\
\frac{d\sigma^V_3}{d\cos \theta} & =  \frac{g_{\chi}^2 \, \tilde{b}_n^2 }{96 \pi s }\frac{12 m_{\chi}^4-4 m_{\chi}^2 t +t^2}{\left(t-m_A^2\right)^2+(\Gamma_{A,3}^V m_A )^2} \frac{ (m_n^2+ m_{\chi}^2-s )^2 + t (s-m_n^2)}{ m_{\chi}^4} \\
\frac{d\sigma^V_4}{d\cos \theta} & =  \frac{g_{\chi}^2 \, \tilde{a}_n^2}{96 \pi s}\frac{12 m_{\chi}^4-4 m_{\chi}^2 t +t^2}{\left(t-m_A^2\right)^2+(\Gamma_{A,4}^V m_A )^2} \frac{m_n^4-2m_n^2(m_{\chi}^2-s)+(m_{\chi}^2- s)^2+s t}{m_{\chi}^4}
\end{align}
\label{eq:Vxsection}
\end{subequations}

We stress that the parameters with tildes have already been introduced in equation \eqref{eq:formfactors}. The decay width $\Gamma_{A,i}^V$ can be easily found using \cite{Berlin:2014tja}.

\begin{figure*}[t]
\begin{center}
\subfloat[\label{sf:V1}]{\includegraphics[scale=0.26]{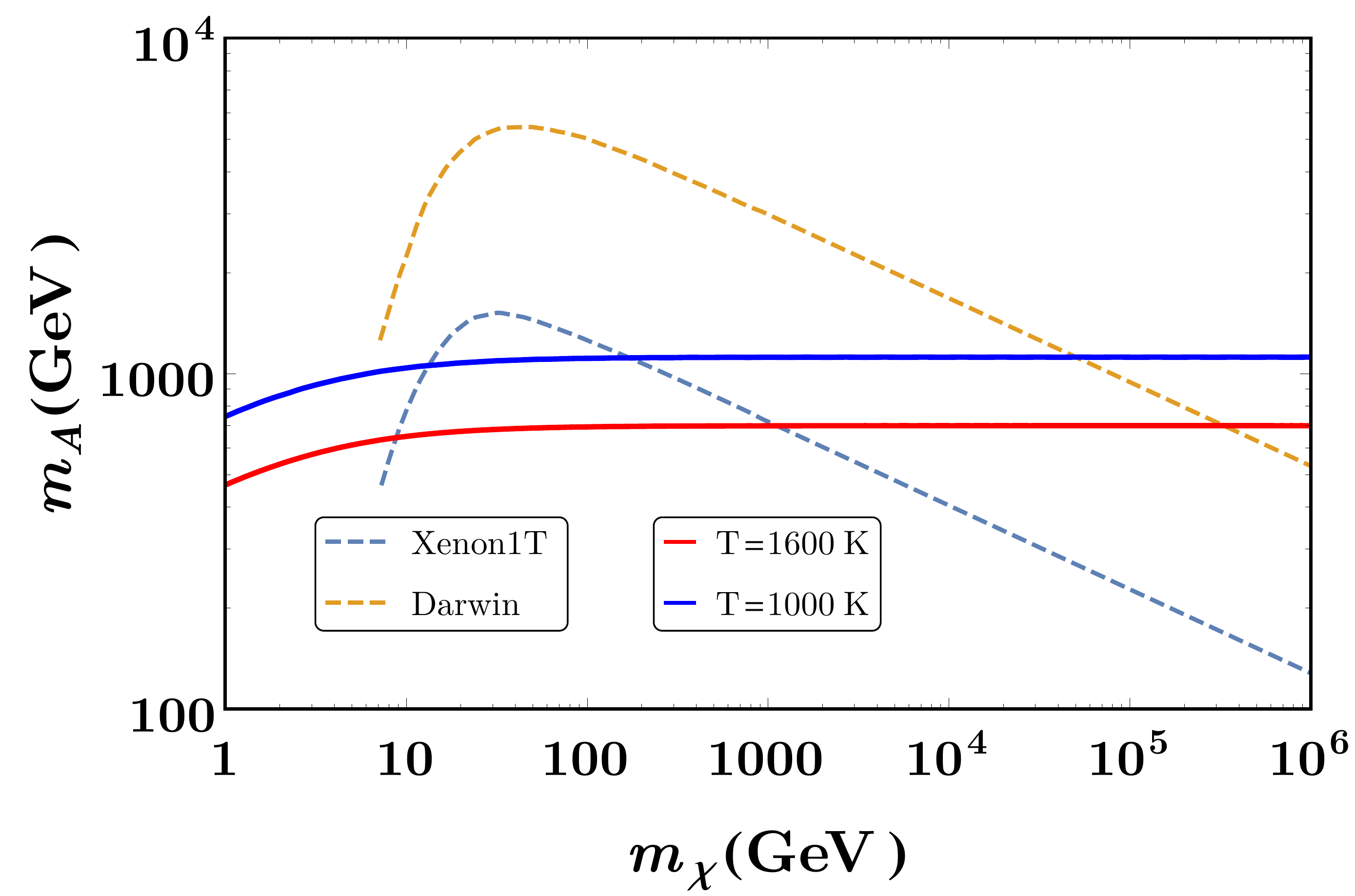}}~~~
\subfloat[\label{sf:V2}]{\includegraphics[scale=0.26]{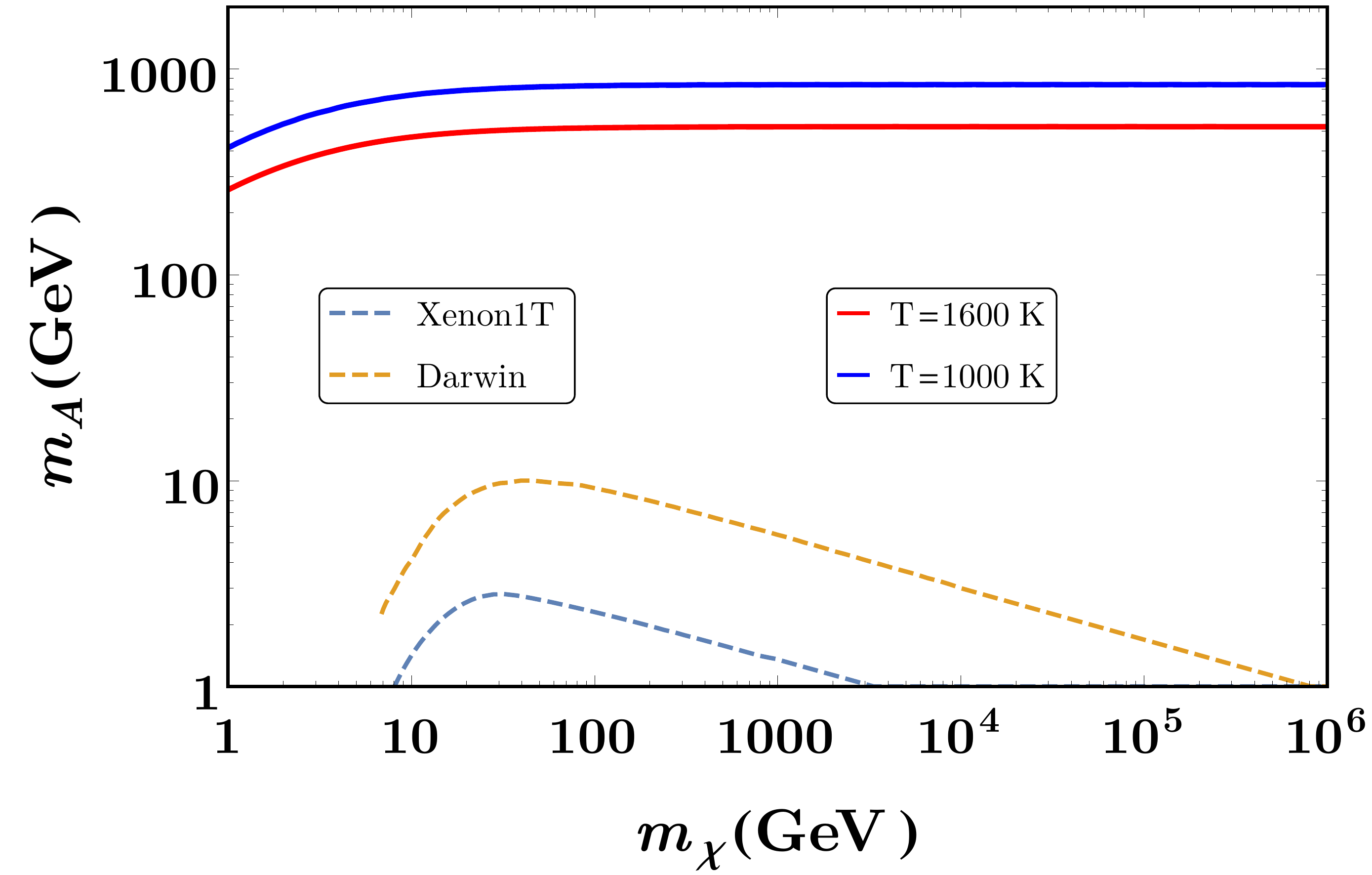}}\\
\subfloat[\label{sf:V3}]{\includegraphics[scale=0.26]{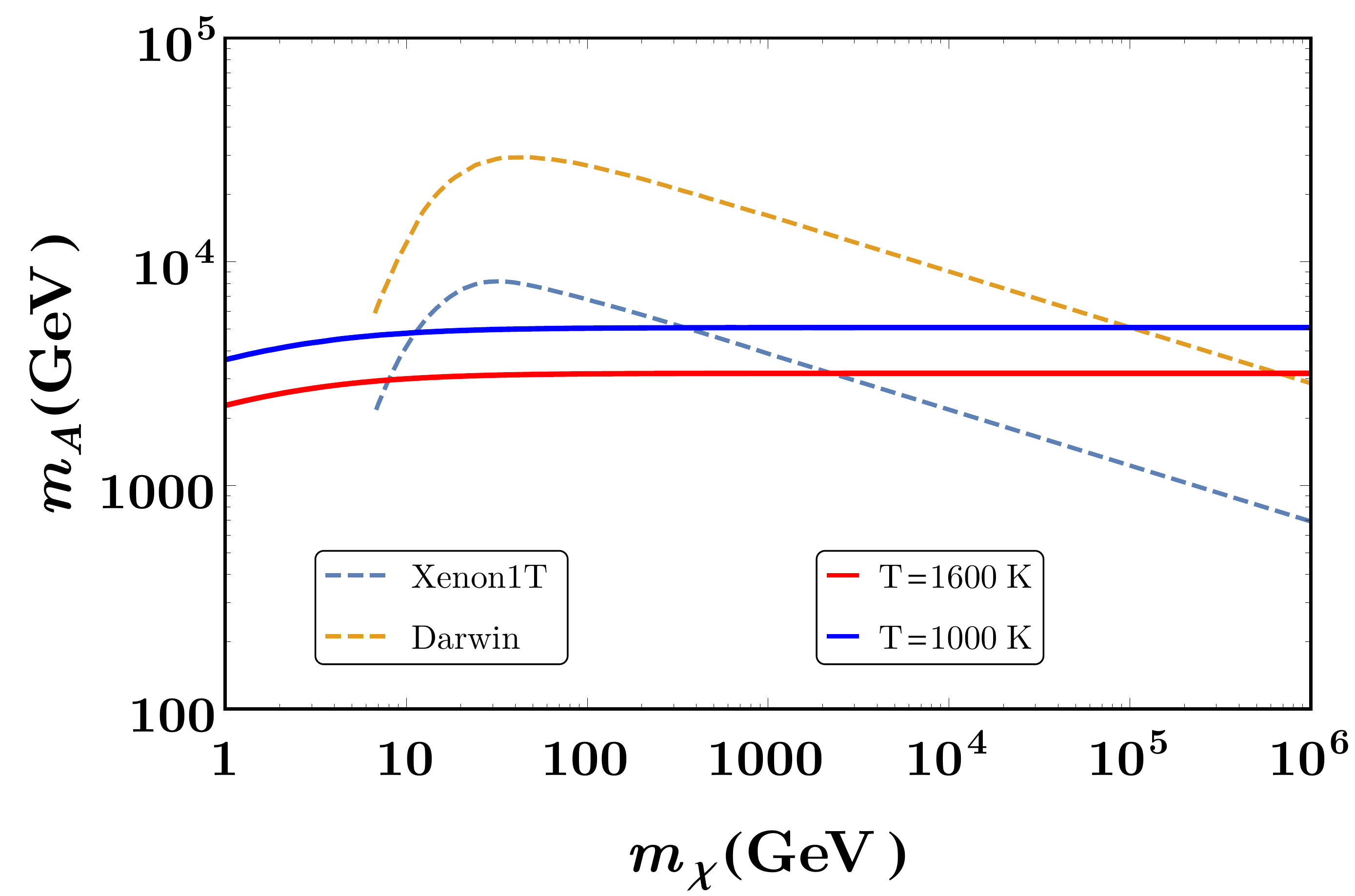}}~~~
\subfloat[\label{sf:V4}]{\includegraphics[scale=0.26]{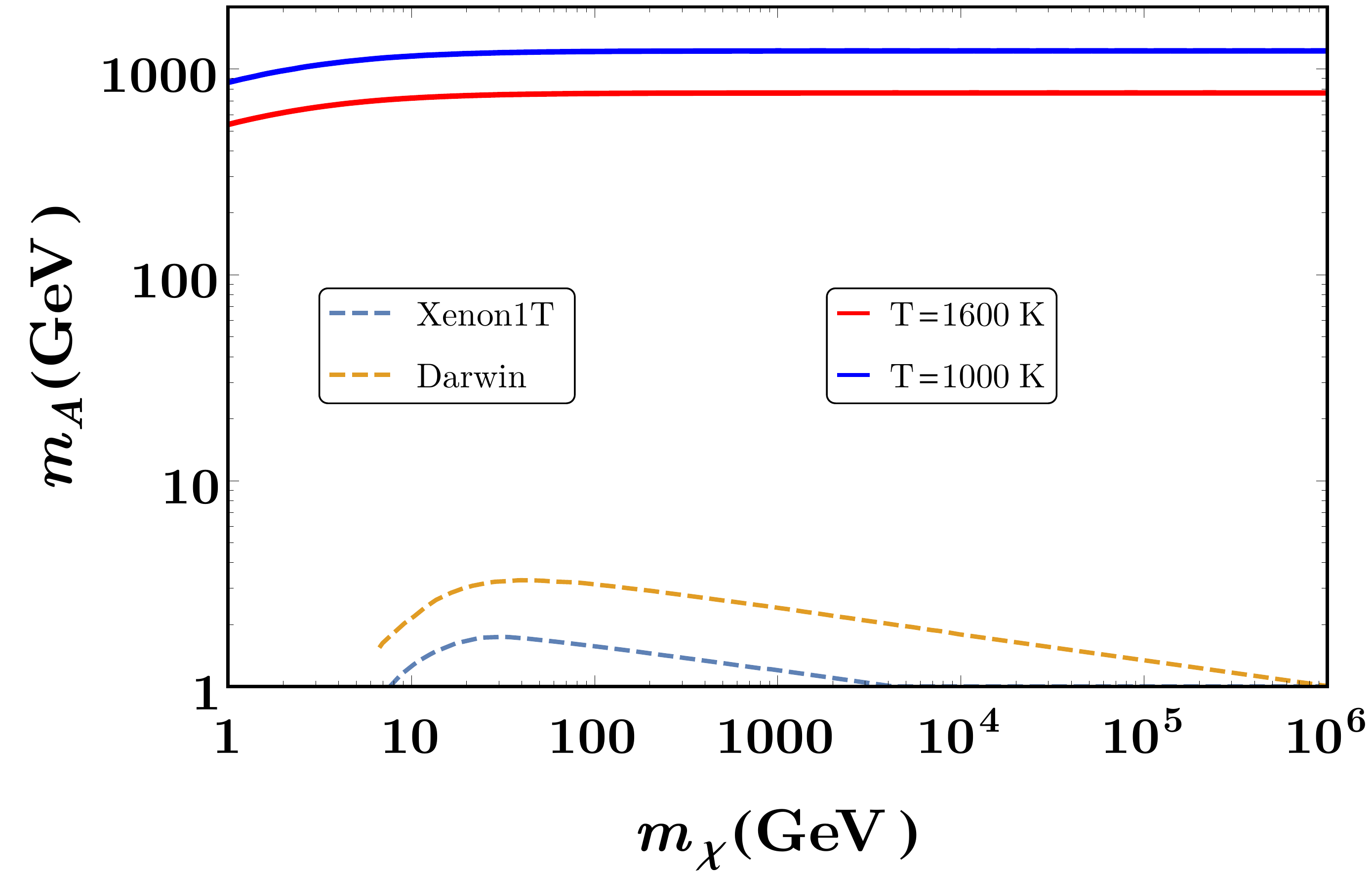}}\\
\caption{Constraint on $m_{\chi}-m_A$ from neutron star heating through DM-neutron elastic scattering. In each panel the red and blue contours correspond to neutron star temperatures $1600\,$ and $1000\,$K, respectively. The complementary bounds from Earth-based direct detection experiments Xenon1T and Darwin are shown by dashed light blue and brown lines. We take $g_i=\mu_{\chi}/m_{\chi}=\lambda_i=0.1$. (a) For the interaction given in equation \eqref{subeq:LV1}. (b) For the interaction given in equation \eqref{subeq:LV2}.(c) For the interaction given in equation \eqref{subeq:LV3}. (d) For the interaction given in equation \eqref{subeq:LV4}.}
\label{fig:NS-DD-V}
\end{center}
\end{figure*}

 We have used equations \eqref{eq:temp} and \eqref{eq:Vxsection} to numerically evaluate the heat in neutron stars due to dark matter interactions.  We emphasize that our reasoning behind constraining DM models with neutron stars is governed by the DM-neutron scattering cross section, which sets capture rate [equation \eqref{eq:f}], and consequently the neutron star temperature [equation \eqref{eq:temp}].

 In the lower and upper panels of figure \ref{fig:NS-DD-V}, we show the limit from neutron star probes in the $m_{\chi}-m_A$ plane, for vector DM interacting with nucleons through scalar and vector mediator, respectively. Our conclusion depends on the  Lorentz structure of the DM-nucleons interaction. Direct detection sets stringent bounds on spin independent operators, conversely, for spin and velocity dependent operators, these constraints are very weak. The heat induced in neutron stars is not affected as much by the presence of velocity dependent operators. For vector DM, similar conclusions are found. In scattering processes that feature velocity suppression, neutron stars can yield bounds that are much stronger than current and existing direct detection experiments. However, when the scattering process is spin independent,  neutron star probes can surpass direct detection bounds in the light and very heavy DM mass regimes.

 \section{Discussions}
 
Our results are applicable for the simplified models described in the previous sections. Variations of these simplified models can impact our conclusions. It is important to mention that our choices for the parameters in the simplifying Lagrangians lead to an important complementarity aspect between direct DM detection and neutron star spectroscopy. These couplings contribute equally in the derivation of the limits rising from direct detection and neutron stars. Hence, a different choice of parameters would change the numerical results, but the qualitative conclusions would remain the same.

Throughout, we have not worried about the region of parameter space that sets the correct relic density, because these simplified models can be embedded in a variety of setups, including non-standard cosmology, which allows one to reproduce the correct relic density in regions of parameter space, originally impossible within the canonical WIMP (weakly interacting massive particle) paradigm. 

As neutron stars, in some cases, have the potential to probe scattering cross sections orders of magnitude below current and future direct detection experiments, neutron stars represent laboratories to probe much smaller DM couplings to fermions. This is important, because if one needs to reside in a corner of parameter space where small couplings are needed, typically that implicates the need for nonstandard cosmology in order to reproduce the correct DM relic density. As nonstandard cosmology studies have grown in interest in light of null results from direct, indirect, and collider searches, neutron stars stand as an exciting avenue to test these scenarios with suppressed couplings.

On a positive note, we have seen complementary studies of DM assessing how much we can narrow down the DM-nucleon scattering cross section combining data from different Earth-based experiments. Neutron stars, which are  subject to a very different type of data, but still rather sensitive to the DM-nucleon scattering cross section might, perhaps significantly, help us to narrow down the properties of the DM particle. This endeavor is a work in progress.

Concerning DM self-capture and evaporation effects, the first can be ignored as long as the DM self-interaction is sufficiently small  \cite{Zentner:2009is,Guver:2012ba}, which typically happens for heavy mediators. The latter can be neglected for DM masses larger than $1$~GeV, as considered throughout this work \cite{Kouvaris:2013awa}.

\section{Conclusions}
\label{sec:conclusion}
In the last few decades, direct detection experiments have made excellent progress pushing down the sensitivity on the DM-nucleon scattering cross section by several orders of magnitude. However, the null results reported thus far set stringent limits. DM-nucleus scattering that grows with atomic mass of the nucleus, or depend on the net spin of the nucleus, are very much constrained by data. However, scattering processes, which feature a momentum dependence, are poorly probed. Nevertheless, the capturing of DM in neutron stars can be rather sensitive to such interactions. We investigate the sensitivity of neutron stars to DM, by taking several simplified models for scalar and vector DM, and put our results into perspective by comparing the sensitivity of current and future direct detection experiments with those from neutron star spectroscopy. 

We have concluded that neutron star spectroscopy offers an orthogonal and complementary probe for DM. This conclusion is readily seen when we compare the limits rising from direct detection experiments with those from neutron star observations for the popular scalar DM model with spin-independent scattering. This setup is well known in the literature because it yields the strongest constraints from direct detection experiments. This scenario is represented in the left upper panel of Figure \ref{fig:NS-DD-S}. Interestingly, the DARWIN projected exclusion region is still surpassed by neutron star temperature measurements for very heavy and light DM. A similar conclusion is faced for vector DM. When the DM-nucleon scattering cross section is momentum suppressed or features a spin-dependence, which is subject to  weaker bounds from direct detection experiments, neutron star spectroscopy is of the utmost importance because they provide limits that are orders of magnitude more stringent. The sensitivity of direct detection experiments to the spin-dependent DM-nucleon scattering is governed by the number of unpaired nucleons. This problem is not present in neutron stars. Furthermore, in the vicinity of neutron stars, DM particles travel at velocities $v \sim 0.3c$, thus overcoming the suppression in scattering rates with momentum dependence. In summary, future observations of neutron stars will become a laboratory for DM discovery, and constitute an orthogonal avenue for DM complementarity studies.

\paragraph*{Acknowledgments\,:} 
TNM thanks Sayan Dasgupta and Rohan Pramanick for their help. TNM acknowledges the IOE-IISc fellowship program for financial assistance. FSQ
is supported by the Sao Paulo Research Foundation (FAPESP) through grant 2015/158971, ICTP-SAIFR FAPESP grant 2016/01343-7, CNPq grants 303817/2018-6 and 421952/2018-0, and the Serrapilheira Institute (grant number
Serra-1912-31613). 


\bibliographystyle{h-physrev}
\bibliography{NS-heating.bib}
\end{document}